\documentclass{article}  
\usepackage{lajolla2006}
\usepackage{graphicx}
\frompage{000} \topage{000}                                              
\def\rightmark{Toward the energy and the system size...}
\def\leftmark{S. Voloshin}

\def\etal{\emph{et al.}}
\newcommand{\PRL}[3]{Phys. Rev. Lett., {\bf {#1}}, {#2} ({#3})}

\newcommand{ \be }{\begin{equation}}
\newcommand{ \ee }{\end{equation}}
\newcommand{ \bea }{\begin{eqnarray}}
\newcommand{ \eea }{\end{eqnarray}}
\newcommand{ \la }{\langle}
\newcommand{ \ra }{\rangle}
\newcommand{ \eps }{\varepsilon}

\newcommand{\mean}[1]{\la {#1} \ra}
\def\v2#1{v_2\{#1\}}

\title{
Toward the energy and the system size dependence of elliptic flow:
working on flow fluctuations
} 
\authors{
{Sergei A. Voloshin
}\\[2.812mm]
{\normalsize
Wayne State University, \\ 
666 W. Hancock, Detroit. 48201, U.S.A.
}}
 
\abstract{
In this talk I concentrate on two topics closely related to the 
understanding of the energy and the system size dependence of the
elliptic flow: determination of the elliptic flow ($v_2$) 
``free'' from the effects of non-flow and flow fluctuations, 
and the role of fluctuations in 
the initial eccentricity of the overlap region. 
I introduce a new approach for the analysis of the distribution in
flow vector, $dP/dq_n$, namely, I propose to use
the Bessel Transform of this ditribution. I show that the Bessel
Transform method is similar to the Lee-Yang Zeroes method, and is
very transparent in its meaning and applications. 
}

 
\begin{document}
 
\maketitle

\setcounter{page}{1}

\section{Introduction}
\label{intro}

Anisotropic flow, and, in particular, elliptic flow, plays a very
important role in our understanding of heavy ion collisions at high
energies.
Since the discovery of the in-plane elliptic flow 
at AGS~\cite{e877-flow}, and later
at CERN SPS~\cite{na49-flow-PRL}, the attention to this phenomena has been
continuously increasing. 
The first RHIC measurements~\cite{star-flow1} by the STAR Collaboration
made it one of the most important observable at 
RHIC~\cite{voloshin-qm2002}. 
The large elliptic flow, along with its
mass dependence at low transverse momenta~\cite{star-flow2,v2pt-mass}
are often used as a strong argument for early thermalization of the
system;
the observed~\cite{star-flow-cqs} constituent quark
scaling~\cite{voloshin-qm2002,voloshin-molnar} 
of elliptic flow strongly suggests that the system spends significant
time in the deconfined state.

In order to understand deeper the physics of the elliptic flow and the
processes governing the evolution of the system, one has to measure
and understand elliptic flow dependence on the system size and
collision energy. 
Having in mind also the dependence on the centrality of the
collision makes the problem indeed 'multi-dimensional'.  
In~\cite{voloshin-poskanzer-PLB} the authors suggested that the
elliptic flow follows a simple scaling in the initial system
eccentricity and the particle density in the transverse plane, 
$v_2/\eps \propto 1/S~dN_{ch}/dy$
(where $\eps$ is the initial system eccentricity and 
$S$ is the area of the overlap region of two nuclei). 
If true, all the data taken at different energies, colliding different
nuclei, and at different centralities should collapse on the same universal
curve. 
Indeed the available data are consistent with such a
scaling~\cite{voloshin-qm2002,na49-flow-PRC}. 
Unfortunately, the systematic uncertainties in the results 
are too large to conclude on how well the scaling holds. 
The main difficulties are the evaluation/elimination  of the
so-called non-flow correlations (azimuthal correlations not related to
the reaction plane orientation), and effects of flow fluctuations, 
along with uncertainties in the calculation of the initial
system eccentricity needed for such a plot. 
 
The non-flow contribution is usually suppressed when one uses
multi-particle correlations to measure anisotropic 
flow~\cite{olli4,star-flow-PRC}.   
Unfortunately, the measurements of flow via correlations of different 
number of particles are also different in their sensitivity to the
event-by-event flow fluctuations\cite{star-flow-PRC,raimond-mike}. 
Thus it become difficult to
disentangle different contributions (non-flow and flow fluctuations)
from measured azimuthal correlations.
It was concluded in~\cite{LYZ-1} that the  Lee-Yang Zeroes
method is the best for suppression of non-flow correlations. 
Initially I thought that this method could be less
sensitive to the flow fluctuations. 
Below I show that this hope was over-optimistic, and 
the results of the Lee-Yang Zeroes
method are affected by flow fluctuations in a similar way as
four-particle cumulant results.

As it was already mentioned, the flow measurements via correlations of
different number of particles, such as two- and four-particle
correlation results, $\v2{2}$ and $\v2{4}$, are
sensitive to flow fluctuations in different ways. 
The contribution of fluctuations should be taken out before one tries to
explore the scaling $v_2/\eps \propto 1/S~dN_{ch}/dy$. 
Here we discuss a different and more convenient approach, namely to
rescale the flow results with an appropriate value of eccentricity that
already include the effect of fluctuations. 
Such type of eccentricities are studied and
presented in the second part of this talk.

\section{Fourier and Bessel Transforms of $dP/dq_x$ and $dP/dq^2$}
\label{BT}

Consider the distribution in $x$ component of $n$-th harmonic flow vector
\be
Q_x=\sum_{i=1}^{M} \cos(n\phi_i) \equiv q_x \sqrt{M},
\ee
where $M$ is the multiplicity of particles used to define the flow vector.
Let us denote the distribution in $Q_x$ in the case of no collective 
flow ($v_n=0$) to be $f_{\rm nf}(x=Q_x)$. In the case of large
multiplicities $M$, this distribution can be approximated by Gaussian,
but we will not need an explicit form here.
In the case of non-zero flow the distribution can be written as
a superposition of ``no-flow'' distributions, $f_{\rm nf}$,  ``shifted'' in the
direction of flow by appropriate amount (for details,
see~\cite{voloshin-zhang}; the necessary condition is $\sqrt{M} \gg 1$):
\be
f(Q_x) \equiv \frac{dP}{dQ_x} 
= \int \frac{d\Psi}{2\pi} f_{\rm nf}(Q_x-  v M\cos(n\Psi)).
\ee
Now let us consider the Fourier transform of this distribution:
\bea
\tilde{f}(k) &=& \la e^{ikQ_x} \ra =
\int dQ_x e^{ikQ_x} \int \frac{d\Psi}{2\pi} f_{\rm nf}(Q_x-vM\cos(n\Psi))
\nonumber \\
&=& 
 \int \frac{d\Psi}{2\pi} \int dQ_x e^{ikQ_x} f_{\rm nf}(Q_x-vM\cos(n\Psi))   
\nonumber \\
&=& 
 \int \frac{d\Psi}{2\pi} e^{ikvM\cos(n\Psi)}   \int dt e^{ikt} f_{\rm nf}(t) 
 = J_0(kvM) \tilde{f}_{\rm nf}(k)
\label{efQx}
\eea
Remarkably, the flow contribution is completely factorized out.
As one expects the function $\tilde{f}_{\rm nf}(k)$ to be close to
Gaussian (at large $M$ it must approach Gaussian  according to the
Central Limit Theorem),
the zeros of the Fourier transform will be determined by
the zeroes of the Bessel function $J_0(kvM)$,  
thus providing a method
of extraction of flow from the distribution in $Q_x$. For example, if $k_1$
is the first zero of the Fourier transform,
\be
v = j_{01} /k_1 M
\ee 
where $j_{01}\approx 2.045$ is the first
 zero of $J_0$ Bessel function. 
The above result is the same as one would obtain applying the
Lee-Yang Zeroes method (using ``sum'' generating
function); in fact the relation to
the Fourier transform was already pointed out in the original paper~\cite{LYZ-1}.

Even more interesting result one obtains considering
two-dimensional distribution, $d^2P/dQ_xdQ_y$. 
In this case
\bea
\tilde{f}(k) &=& 
\int dQ_x e^{ik_x Q_x} dQ_y e^{ik_y Q_y}  \frac{d^2P}{dQ_x dQ_y}
\nonumber \\
&=& 
\int dQ J_0(kQ) \frac{dP}{dQ}
~\sim~
 J_0(kvM), 
\eea
which, as shown above, is reduced to the Bessel transform of the
distribution in the magnitude of the flow vector. 
Note that in this approach (valid in the limit of $\sqrt{M} \gg 1$) the
 flow contribution is completely decoupled from all other
 correlations. It happens due to the collective nature of flow.
Note also that in the same limit one expects the distribution in flow
 vector to be Gaussian due to the Central Limit Theorem, thus
 explaining why the fitting of the distribution to the form derived
 in~\cite{voloshin-zhang} (such fits have been
 used in~\cite{e877-flow,star-flow-PRC}) 
is also non-sensitive to non-flow correlations. Thus
 in this limit all three methods, the Bessel Transform, Lee-Yang
 Zeroes, and fitting $q$-distribution become very similar, if not equivalent.

Fig.~1 (left panel) shows the fit to the $q$-distribution of the simulated data (four
million events with 400 particles in each event; 50\% events have flow
$v=0.04$ and 50\% with $v=0.06$).  Right panel shows the Bessel
transform of the same distribution. One finds that the results
obtained with two methods are indeed very close to each other. For
a discussion of a small deviation from 0.05, expected for this case,
see below.

\begin{figure}[htb]
\vspace*{-0.5cm}
                 \includegraphics[width=0.52\textwidth]{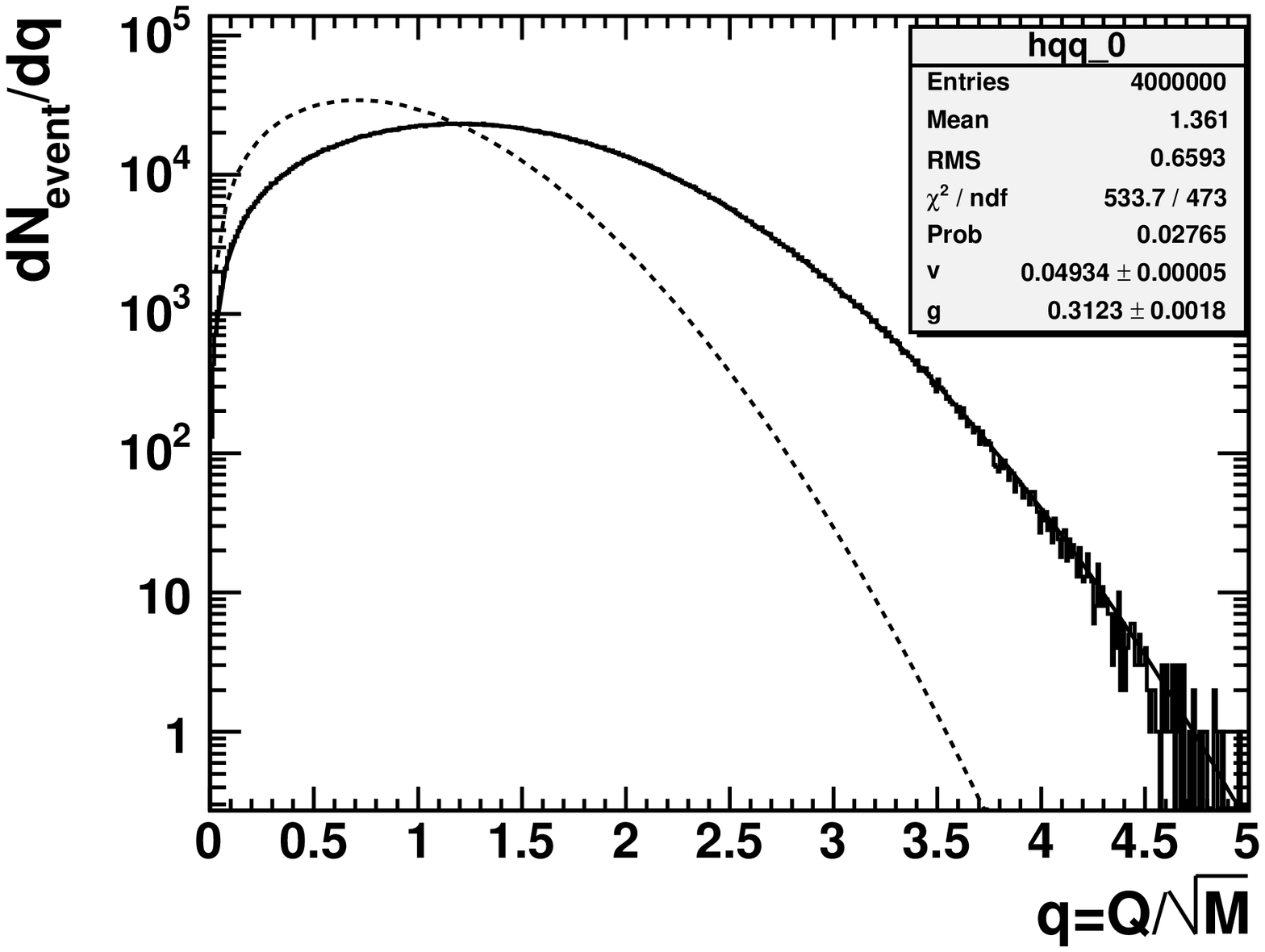}
                 \includegraphics[width=0.52\textwidth]{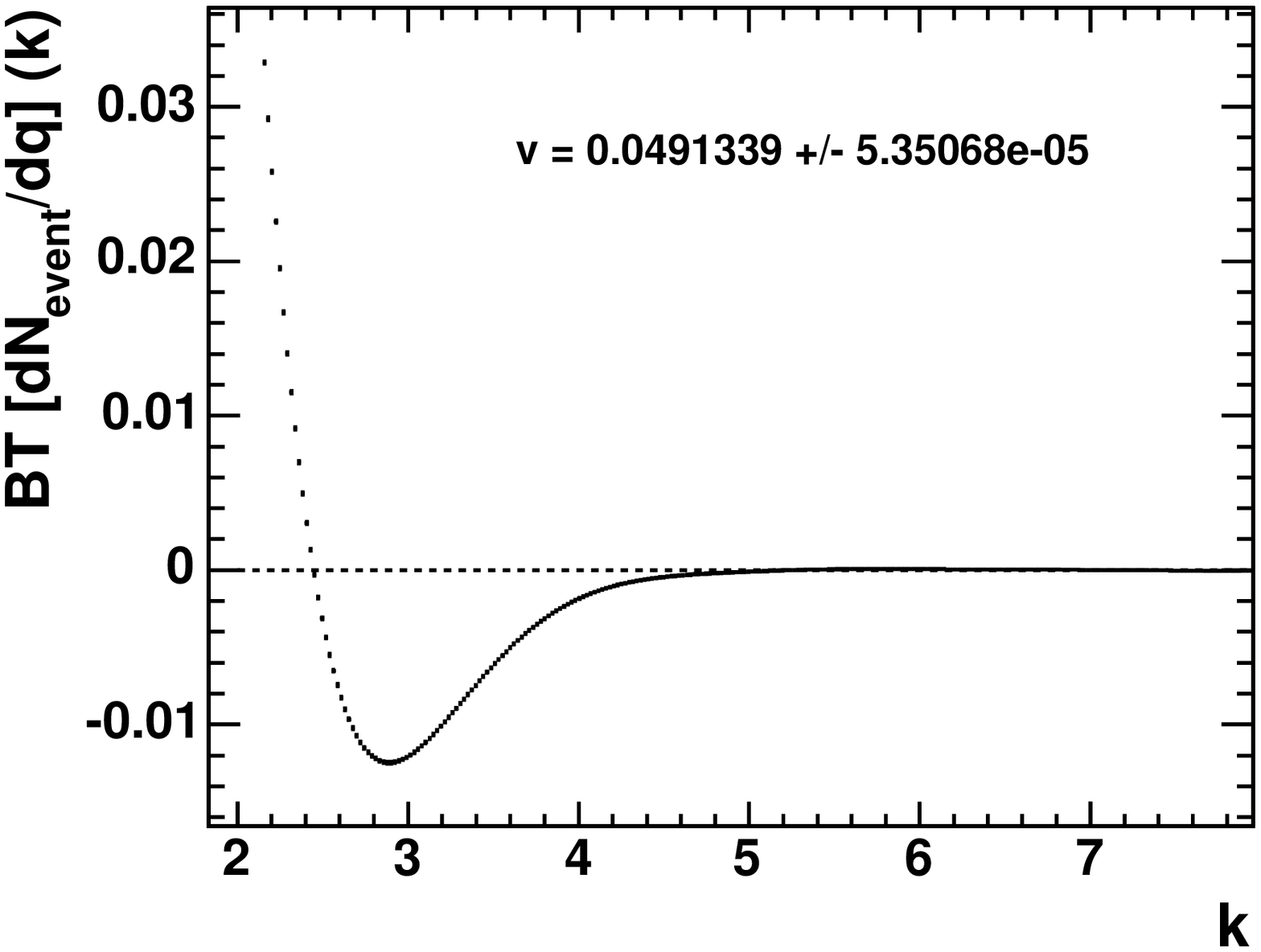}
\vspace*{-1.cm}
\caption[]{
(left panel) $q$-distribution for simulated events shown together with
fit function. Dashed line shows the expected distribution in case of $v=0$.
(right panel) Bessel transform of the distribution on the left. (See
the last row in Table~1 for the flow values obtained in this case).
\vspace*{-0.3cm}
} \label{fig1}
\end{figure}

Though our goal here is to calculate the integrated flow, it is
straight-forward to generalize the method for the calculation of
differential flow as well.
For that one considers 
\bea
\la \cos(\phi_i) \, e^{ikQ_x} \ra = 
\int \frac{d\Psi}{2\pi} d\phi_i (1+2v\cos(\phi_i-\Psi)) \cos(\phi)
 e^{ikQ_x} f(Q_x)
\nonumber \\
=
\int \frac{d\Psi}{2\pi} v \cos(\Psi)  e^{ik \hat{v} M\cos(\Psi)}
\tilde{f}_{\rm nf}(k)
~=~ v J_1(k \hat{v} M)\tilde{f}_{\rm nf}(k),
\eea
where we have introduced notation $\hat{v}$ for the average flow in
the region of the flow vector definition. 
From the above equation one can obtain the flow $v$ by different
methods. The one suggested in~\cite{LYZ-1}
is based on the relation 
\be
\la Q_x e^{ikQ_x} \ra = \hat{v} M  J_1(k \hat{v} M)\tilde{f}_{\rm nf}(k).
\ee
Combining Eqns. 6 and 7 one finds:
\be
v = \hat{v} M \frac{\la \cos(\phi_i) e^{ikQ_x} \ra}{\la Q_x e^{ikQ_x}
  \ra}.
\ee
Alternative way would be to use the expression \ref{efQx} directly. Then
\be
v = \frac{\la \cos(\phi_i) e^{ikQ_x} \ra} {\la e^{ikQ_x} \ra}
\frac{J_0(k\hat{v}M)} {J_1(k\hat{v}M)}
\ee
More detailed discussion of the differential flow measurements using
Bessel Transform is deferred for later publication.

The main advantage of using the Bessel Transform (as well as Lee-Yang
Zeroes method, or fitting the $Q$-distribution, as all three methods
are very close in their assumptions and limitations) is that it is much less
sensitive to the non-flow contribution compared to the standard
method~\cite{method}. 
The sensitivity of these methods to flow fluctuations is not that clear
from the above equations. 
As it follows from the simulation results presented in Table~1, 
the actual sensitivity to fluctuations is very similar to that of
four-particle cumulant approach.

\begin{table}[hb] 
\vspace*{-12pt}
\caption[]{Flow values (in \%) obtained by different methods. 
}\label{tab1}
\vspace*{-14pt}
\begin{center}
\begin{tabular}{lllll}
\hline\\[-10pt]
                 & $v_2\{2\}$ & $ v_2\{4\}$ & $v_2$\{BT\} & $v_2$\{q-dist\}\\ 
\hline\\[-10pt]
without non-flow \\
\hline\\[-10pt]

100\% $v=6.0$   & 6.005$\pm$0.003 & 6.006$\pm$0.006 & 6.018$\pm$0.004
		& 5.999$\pm$0.002 \\

100\% $v=4.0$	& 3.995$\pm$0.004 & 3.977$\pm$0.002 & 4.013$\pm$0.008 & 3.997$\pm$0.003\\

50/50 $v=4$ \& $v=6$ & 5.100$\pm$0.004 & 4.895$\pm$0.007 &
                 4.908$\pm$0.004 & 4.920$\pm$0.004 \\

\hline\\[-10pt]
including non-flow\\
\hline\\[-10pt]

100\% v=6.0	  	& 6.495$\pm$0.003 & 6.005$\pm$0.007 &
                 6.028$\pm$0.004 & 6.033$\pm$0.004 \\

100\% v=4.0		& 4.703$\pm$0.004 & 4.013$\pm$0.002 &
                 4.011$\pm$0.002 & 4.038$\pm$0.009 \\

50/50 v=4 \& v=6 & 5.670$\pm$0.005 & 4.905$\pm$0.008 &
                 4.913$\pm$0.005 & 4.933$\pm$0.005 \\

\hline 
\end{tabular}
\end{center}
\end{table}

Table~1 present the results obtained by different methods on simulated
data without flow fluctuations, 100\% events with $v=0.04$ or
$v=0.06$, and 50/50 mixture of the above. Last three rows present
results from simulations including large non-flow correlations
(simulated by pairs of particles with erectly the same azimuth). 
The following observations follow: $v\{2\}$ results, as expected are
sensitive to both non-flow effects and flow fluctuations (the presented
results coincide within statistical errors with analytical expectations).
$ v_2\{4\}$, $v_2$\{BT\}, and $v_2$\{q-dist\} are not affected by
non-flow. For the case of 50/50 mixture of events with different flow
values, $ v_2\{4\}$ yields value below 0.05 due to flow fluctuations.
What was somewhat unexpected is that the sensitivity (bias) of
$v_2$\{BT\}, and $v_2$\{q-dist\} to flow fluctuations is very
similar to that of $ v_2\{4\}$. More detailed examination of the
Eq.~3 and expansion of the Bessel function around $j_{01}$ up to the
second order confirms this observation made on simulated data.
This is also illustrated in Fig.~2 (read caption).  

\begin{figure}[htb]
\vspace*{-0.3cm}
                 \includegraphics[width=0.62\textwidth]{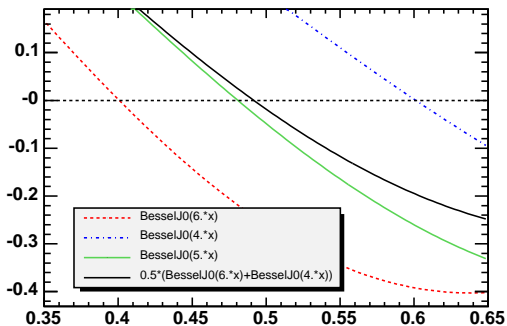}
\vspace*{-.8cm}
\caption[]{
The most left and most right curves shows the expectations for Bessel
Transform of $q-$distribution for 4\% and 6\% flow. The solid black
line shows the average of these two. The green line shows the expectations
for 5\% flow. The difference in zeroes of black and green lines
indicates the effect of flow fluctuations.
\vspace*{-0.5cm}
} \label{fig2}
\end{figure}

\section{Eccentricity and eccentricity fluctuations}
\label{epsilon}

If the initial conditions are totally isotropic -- no anisotropic flow 
is possible. So one concludes that at small initial anisotropies 
the final anisotropic flow must be {\em linearly} proportional 
to the initial spatial anisotropy of the system. 
Although it is not clear what one should use
(what is most relevant) for the higher harmonic flow, in case of
 the elliptic flow 
it is relatively straight-forward: one should come up with some measure 
that differentiate the system geometry along the impact parameter, 
in-plane, and, correspondingly, out-of-plane, such as
$(R_y-R_x)/(R_y+R_x)$ or $(R_y^2-R_x^2)/(R_y^2+R_x^2)$. 
where radii $R_x$ and $R_y$ are some characteristics of the system
in- and out-of- the reaction plane directions. 
Note, that different definitions of the eccentricity give different
values, but this is not a problem if one uses the {\em same} 
definition for all systems/centralities/experiments used in a comparison, and
as long as the eccentricity itself is small. 
Recall that the goal of this exercise is to 
exploit the fact that at small eccentricities the final flow should be 
proportional to eccentricity. 
(Note that such proportionality would not at all be trivial if eccentricity is large). 
Having said all that, the most often used definition of eccentricity is
\be
\eps \equiv \frac{\mean{y^2-x^2}}{\mean{y^2+x^2}},
\ee
where the $x$ direction is taken along the direction of the impact
parameter,
and the average, with some weight, is taken over all points 
of the system. 
For the weight one can use the participating nucleon or produced particle
densities,
entropy or energy densities, density of binary collisions, etc..  
A priory it is not known what weight would be
the best to use, but again, note that at small eccentricities it
should not matter provided  the {\em same} weight is used for all the data.

In the standard {\em optical Glauber} definition of the eccentricity
the weight function is assumed to be the same for all events
with the same impact parameter; any event-by-event fluctuations are excluded.
In reality, even at fixed impact parameter, 
the eccentricity could fluctuate for many different reasons:
the number of individual nucleons (quarks, partons) participating 
in the collision could
fluctuate, their position in the transverse plane also fluctuate, the
multiplicity of produced particles in each of ``individual collision''
could fluctuate along with deposited energy, etc..
The first serious attempt to calculate the role of such kind of
fluctuations for flow analysis
in {\em Monte-Carlo Glauber} model was done by 
Snellings and Miller~\cite{raimond-mike}
who calculate the fluctuation in 
\be
\eps_{std}=\frac{\mean{y^2-x^2}}{\mean{y^2+x^2}},
\ee
with ${x,y}$ being the position of the participating nucleons.
It was also soon realized, that to take the fluctuations fully into
account one should take into account also fluctuations in the
direction of the major axes of the overlap region as well as the
fluctuation in its center of gravity~\cite{manly-qm2005}:
\be
\eps_{part}=\frac{\mean{y'^2-x'^2}}{\mean{y'^2+x'^2}},
\ee
where the eccentricity is calculated relative to the new coordinate
system defined by the major axis of the initial system region.
The results of the corresponding calculations are presented in
Fig.~3. 
Note significantly larger difference between the standard and participant
eccentricities in Cu+Cu collisions compared to Au+Au collisions.
Using $\eps_{part}$ instead of $\eps_{std}$ significantly improves
scaling of $v_2/\eps$~\cite{manly-qm2005}.

With elliptic flow assumed to be proportional to the eccentricity in an
event, $v_2 \propto \eps$, flow fluctuations should be determined (at
least partially) by the fluctuation in eccentricity. Thus, 
$v_2\{2\} \propto \eps\{2\}$ and $v_2\{4\} \propto \eps\{4\}$, where 
\be
 (\eps\{2\})^2  \equiv  \mean{\eps^2};\;\; 
 (\eps\{4\})^4  \equiv  2 (\eps\{2\})^4 -
\mean{\eps^4}.
\ee
The results for $\eps\{2\}$ and $\eps\{4\}$ are also presented in Fig.~3.
Two remarks are in order for this plot. The ratio of $\eps\{2\}$'s for Cu+Cu
and Au+Au collisions is larger than unity, which, is used, would improve the
scaling observed in~\cite{manly-qm2005}.
Very small values of
$\eps\{4\}$ for mid-central and central collisions could complicate
the analysis of $v_2\{4\} / \eps\{4\}$ which could become very
sensitive to the exact scale of flow fluctuations.

\begin{figure}[htb]
\vspace*{-0.2cm}
                 \includegraphics[width=0.52\textwidth]{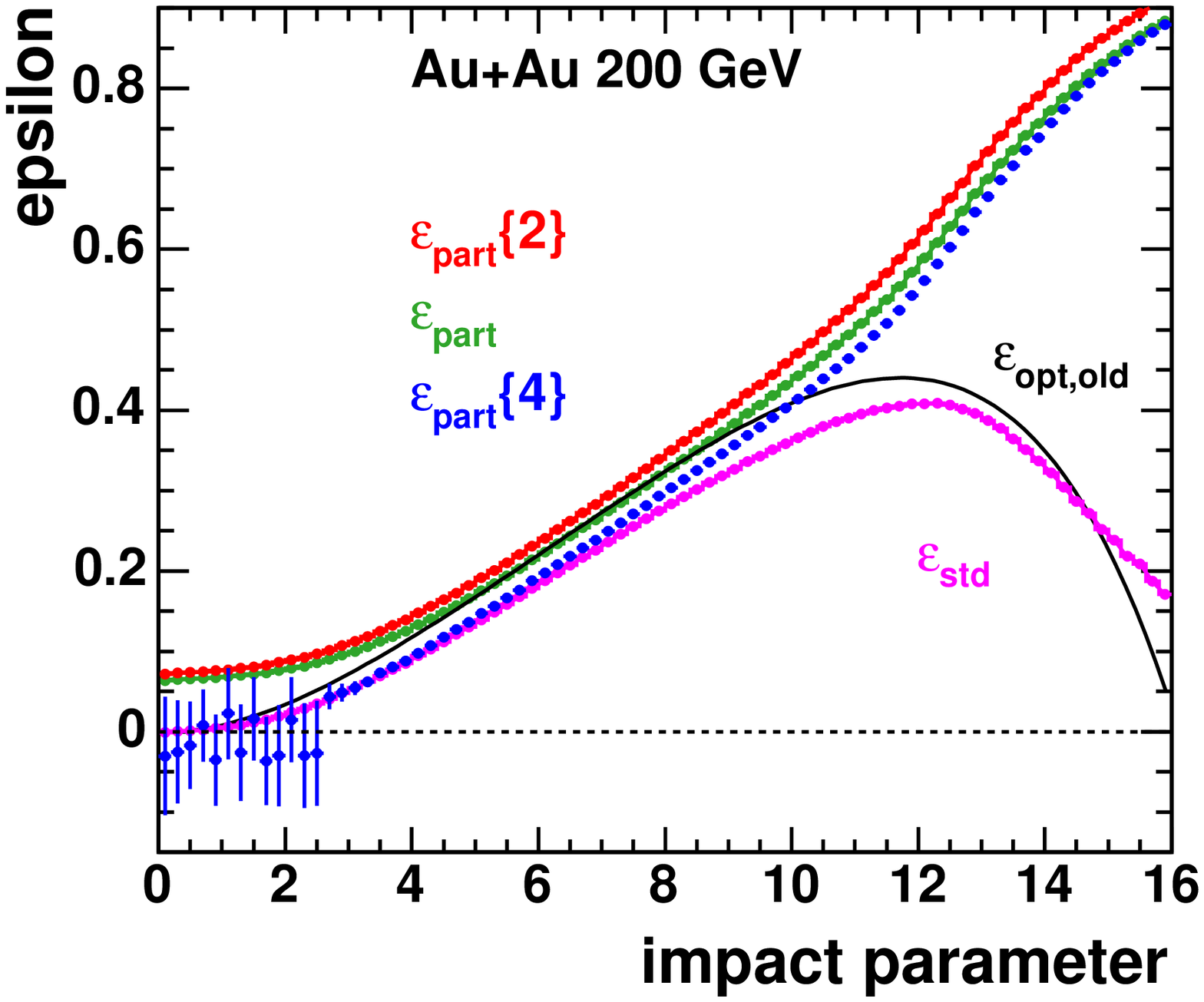}
                 \includegraphics[width=0.52\textwidth]{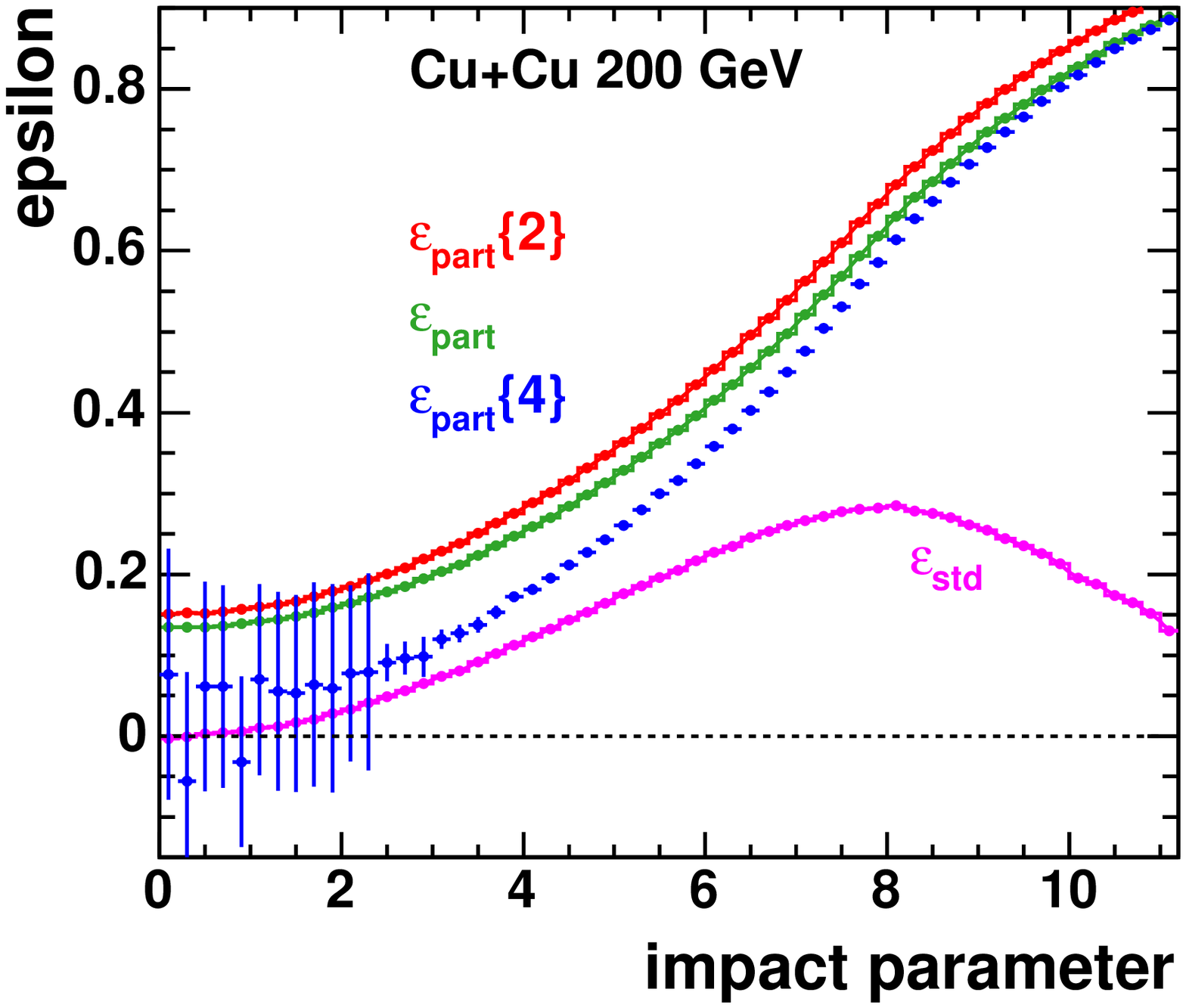}
\vspace*{-1.cm}
\caption[]{
Eccentricity values for Au+Au and Cu+Cu collisions at
$\sqrt{s_{NN}}=200$~GeV as function of
impact parameter. For comparison, the solid line in the left panel shows the results of
optical Glauber calculations used in~\cite{star-flow-PRC,na49-flow-PRC}.
\vspace*{-0.5cm}
} \label{fig3}
\end{figure}

\section{Conclusions}\label{concl}
A new method of flow analysis, the Bessel transform of $q$-distribution
has been introduced. It is shown that this method is very similar to
the Lee-Yang Zeroes (using {\em sum} generating function) 
and the fitting of $q$-distribution methods. It was
also shown that its sensitivity to flow fluctuations (as well as that
of the Lee-Yang Zeroes method) is similar to the four particle cumulant results.

Eccentricity fluctuations relevant for flow measurements $v\{2\}$ and  $v\{4\}$
are presented for Au+Au and Cu+Cu collisions at RHIC energies.

\section*{Acknowledgments}
I thank the organizers for a very interesting and stimulating  workshop. I
also thank the members of the STAR flow group for numerous and fruitful
discussions of different results presented in this talk. Monte-Carlo
Glauber simulations have been performed by J. Gonzales (UCLA). 
Financial support provided by US Department of Energy Grant No. DE-FG02-92ER40713.

\vfill\eject

\begin{thebibliography}{9}  
  
\bibitem{e877-flow}
  J.~Barrette {\it et al.}  [E877 Collaboration],
  Phys.\ Rev.\ Lett.\  {\bf 73}, 2532 (1994);
  J.~Barrette {\it et al.}  [E877 Collaboration],
  Phys.\ Rev.\ C {\bf 55}, 1420 (1997).

\bibitem{na49-flow-PRL}
  H.~Appelshauser {\it et al.}  [NA49 Collaboration],
  Phys.\ Rev.\ Lett.\  {\bf 80}, 4136 (1998)

\bibitem{star-flow1}
	K.H. Ackermann \etal [STAR Collaboration], \PRL{86}{402}{2001}. 

\bibitem{voloshin-qm2002}
  S.~A.~Voloshin,
  Nucl.\ Phys.\ A {\bf 715}, 379 (2003)

\bibitem{star-flow2}
  C.~Adler {\it et al.}  [STAR Collaboration],
  Phys.\ Rev.\ Lett.\  {\bf 87}, 182301 (2001).

\bibitem{v2pt-mass}
  P.~Huovinen, P.~F.~Kolb, U.~W.~Heinz, P.~V.~Ruuskanen and S.~A.~Voloshin,
  Phys.\ Lett.\ B {\bf 503}, 58 (2001);
  S.~A.~Voloshin,
  Phys.\ Rev.\ C {\bf 55}, 1630 (1997).

\bibitem{star-flow-cqs}
  J.~Adams {\it et al.}  [STAR Collaboration],
  Phys.\ Rev.\ Lett.\  {\bf 92}, 052302 (2004).
 
\bibitem{voloshin-molnar}
  D.~Molnar and S.~A.~Voloshin,
  Phys.\ Rev.\ Lett.\  {\bf 91}, 092301 (2003).

\bibitem{voloshin-poskanzer-PLB}
  S.~A.~Voloshin and A.~M.~Poskanzer,
  Phys.\ Lett.\ B {\bf 474}, 27 (2000).

\bibitem{na49-flow-PRC}
  C.~Alt {\it et al.}  [NA49 Collaboration],
  Phys.\ Rev.\ C {\bf 68}, 034903 (2003).

\bibitem{olli4}
  N.~Borghini, P.~M. Dinh, and J.~Y.~Ollitrault,
  Phys.\ Rev.\ C {\bf 63}, 054906 (2001).

\bibitem{star-flow-PRC} 
	STAR Collaboration, C. Adler et al., 
	Phys. Rev. C {\bf 66}, 034904 (2002).

\bibitem{raimond-mike}
  M.~Miller and R.~Snellings,
  arXiv:nucl-ex/0312008.

\bibitem{LYZ-1}
  R.~S.~Bhalerao, N.~Borghini and J.~Y.~Ollitrault,
  Nucl.\ Phys.\ A {\bf 727}, 373 (2003).
  
\bibitem{method}
        A.M.~Poskanzer and S.A.~Voloshin,
        Phys. Rev. C {\bf 58}, 1671, 1998.

\bibitem{voloshin-zhang}
  S.~Voloshin and Y.~Zhang,
  Z.\ Phys.\ C {\bf 70}, 665 (1996)

\bibitem{manly-qm2005}
  S.~Manly {\it et al.}  [PHOBOS Collaboration],
  arXiv:nucl-ex/0510031.


\end{thebibliography}
\end{document}